\documentclass[12pt]{article}
\begin{document}
\title{\bf Trace formulas and spectral statistics of\\ 
diffractive systems}
\author{{\it E. Bogomolny}\\
Laboratoire de Physique Th\'eorique et\\
Model\`es Statistiques\\
Universit\'e de Paris XI, B\^at. 100\\
91405 Orsay Cedex, France}

\maketitle
 
\begin{abstract}
Diffractive systems are quantum-mechanical models with point-like
singularities where usual semiclassical approximation breaks down. An 
overview of recent investigations of such systems is presented. The following
examples are considered in details:
(i) billiards (both integrable and chaotic) with small-size scatterers, (ii)
pseudo-integrable polygonal plane billiards, and (iii) billiards with the
Bohr-Aharonov flux lines. First, the diffractive trace formulas are discussed 
with particular emphasis on models where the diffractive coefficient
diverges in certain directions. Second, it is demonstrated that  the spectral
statistics of diffractive models are different from the statistics of both 
integrable and chaotic systems. The main part of the paper is devoted to
analytical calculations of spectral statistics for certain diffractive
models. 
\end{abstract}

\pagebreak

\section{Introduction}
Quantum chaos (or chaology) is a part of theoretical physics whose aim is
the investigation of (multi-dimensional) quantum problems in the
semiclassical limit $\hbar \rightarrow 0$. The modern development of this
field is based on the semiclassical representation of the Green function,
$G(\vec{x}, \vec{x}\ ')$, through the sum over classical trajectories
connecting points $\vec{x}$ and $\vec{x}\ '$ (see e.g. \cite{1})
\begin{equation}
G(\vec{x}, \vec{x}\ ')=\sum_{cl.\ tr.}A_{tr}(\vec{x}, \vec{x}\ ')
\exp (\frac{i}{\hbar}S_{cl}(\vec{x}, \vec{x}\ '))
\label{1}
\end{equation}
where the action, $S_{cl}(\vec{x}, \vec{x}\ ')$, and the pre-factor,
$A_{tr}(\vec{x}, \vec{x}\ ')$ are computed from pure classical mechanics.
In such an approach one implicitly assumes that the classical limit of quantum
mechanical problem does exist which is not always the case. For a long time
(see e.g. \cite{2}) it was known that there are quantum systems without
`good' classical limit. 

This paper is devoted to the investigation of a particular type of such
models, namely, diffractive systems whose characteristic property is the
presence of point-like singularities (real or effective) in the quantum
Hamiltonian. The existence of singularities in classical mechanics leads
to the impossibility of  continuing classical trajectories which hit these
singularities. In quantum mechanics the situation is less dramatic
\cite{2}-\cite{6}. Each point-like singularity can be described by  a
diffraction coefficient, $D(\vec{n}, \vec{n}')$, which determines the
scattering amplitude on the singularity. We normalize it  in such a way
that the semiclassical expansion of the Green function in the whole space in
the presence of a singularity at point $\vec{x}_0$ has the following form
\begin{equation} 
G(\vec{x}, \vec{x}\ ')=G_0(\vec{x}, \vec{x}\ ')+\frac{\hbar^2}{2m}
\sum_{\vec{n},\vec{n}'}G_0(\vec{x}, (\vec{x}_0,\vec{n})) D(\vec{n}, \vec{n}')
G_0((\vec{x}_0, \vec{n}'),\vec{x}\ ')
\label{2}
\end{equation}
where $G_0(\vec{x}, \vec{x}\ ')$ is the Green function (\ref{1}) in the
absence of singularity and $G_0(\vec{x}, (\vec{x}_0,\vec{n}))$ is a term in
(\ref{1}) corresponding to a classical trajectory starting from  point
$\vec{x}$ and ending at point $\vec{x}_0$ with momentum in the direction
$\vec{n}$.

The knowledge of the Green function permits to write down the trace formula
for diffractive systems \cite{3,4}. In the simplest case, when only one
singularity is present, the density of states can be written as a sum of
three terms
\begin{equation}
d(E)=\bar{d}(E)+d_p(E)+d_d(E)
\label{2b}
\end{equation}
where $\bar{d}(E)$ is the smooth part of the level density given by the
Thomas-Fermi term (plus corrections if necessary) \cite{1}, $d_p(E)$ is the
contribution from classical periodic orbits, and $d_d(E)$ is the diffractive
contribution.

For integrable billiards \cite{2b}
\begin{equation}
d_p(E)=\frac{A}{2\pi}\sum_{ppo}\sum_{n=1}^{\infty}
   \frac{1}{\sqrt{k nl_p}}\cos(knl_p-\frac{\pi}{4}-\frac{\pi}{2}n\nu_p).  
\label{2c}
\end{equation}

For chaotic billiard systems \cite{1}
\begin{equation}
d_p(E)=\sum_{ppo}\frac{l_p}{\pi k}\sum_{n=1}^{\infty}\frac{1}{|\det
  (M^n_p-1)|^{1/2}}\cos (knl_p-\frac{\pi}{2}n\nu_p).
\label{2d}
\end{equation}
In these formulas the summation is performed over all primitive periodic
orbits (ppo) and their repetitions, $l_p$ is the length of the ppo, $\nu_p$ is
its Maslov index, and $M_p$ is the monodromy matrix of the ppo.

$d_d(E)$ in Eq.~(\ref{2b}) is the contribution from classical orbits (called
later diffractive orbits) which start from the singularity and end at it. 
\begin{eqnarray}
d_d(E)&=&\sum_{m=1}^{\infty}\frac{1}{\pi m}\frac{\partial }{\partial E}
\mbox{ Im}\sum_{\vec{n},\vec{n}'}
G(\vec{n}_1,\vec{n}_1')D(\vec{n}_1',\vec{n}_2)G(\vec{n}_2,\vec{n}_2')D(\vec{n}_2',\vec{n}_3)
\nonumber \\
& &\ldots G(\vec{n}_m,\vec{n}_m')D(\vec{n}_m',\vec{n}_1).
\label{3}
\end{eqnarray}
Here $G(\vec{n},\vec{n}')$ is the contribution to the Green function
(\ref{1}) from a diffractive orbit which starts from the singular point
with momentum in  direction $\vec{n}$ and ends at this point with 
momentum in  direction $\vec{n}'$. The sums in (\ref{3}) can be
transformed to
\begin{equation}
d_p(E)=-\frac{1}{\pi} \mbox{ Im }\frac{\partial }{\partial E}\ln \det
(1-\hat{K})
\label{4}
\end{equation}
where the operator $\hat{K}$ is defined as follows
\begin{equation}
(\hat{K} \phi)(\vec{n})=\sum_{\vec{n}',\vec{n}''} G(\vec{n},\vec{n}')
D(\vec{n}',\vec{n}'')\phi(\vec{n}'').
\label{5}
\end{equation}
Eqs.~(\ref{4})-(\ref{5}) lead to  a formal `quantization' condition of
diffractive systems (from which energy levels can be computed)  
\begin{equation}
\det (1-\hat{K})=0.
\label{6}
\end{equation}
The main purpose of this paper is to discuss statistical properties of
energy levels (i.e. the spectral statistics) of
diffractive systems. The plan of the paper is the following. In Section 2 
integrable systems with a diffractive center are considered and it is
demonstrated that the spectral statistics of these systems can be computed
analytically. In Section 3 the pseudo-integrable polygonal billiards and in
Section 4 the billiards with an Aharonov-Bohm flux line are investigated.
These models are examples of diffractive systems where the diffractive
coefficient diverges in certain directions which considerably complicates
all calculations. First, the trace formulas for  these systems are derived
and it is demonstrated that they differ from Eq.~(\ref{3}). Second,  
by combining numerical and analytical arguments we investigate their spectral 
statistics. Finally, in Section 5 it is proved that the addition of a
diffractive center to chaotic models does not change its spectral statistics.

\section{ Integrable models with diffractive center}

The simplest example of diffractive systems with a constant diffractive
coefficient has been proposed in \cite{7} and it consists of an integrable
model (below we shall consider a 2-dimensional rectangular billiard) with a
$\delta$-function potential
\begin{equation}
V(\vec{x})=\lambda \delta(\vec{x}-\vec{x}_0).
\label{7}
\end{equation}
The quantization condition (\ref{6}) in this case takes a particular simple
form
\begin{equation}
  \lambda G(\vec{x}_0,\vec{x}_0)=1
\label{8}
\end{equation}
or (ignoring renormalization problems \cite{8}) 
\begin{equation}
  \lambda \sum_{n=1}^N\frac{|\psi_n(\vec{x}_0)|^2}{E-e_n}=1
\label{9}
\end{equation}
where $e_n$ and $\psi_n(\vec{x})$ are eigenvalues and eigenfunctions of
the unperturbed system (i.e. without the $\delta$-function potential 
(\ref{7})).
The natural question  arises: what is the distribution of the new eigenvalues,
$E_n$, provided the distributions of $e_n$ and $\psi_n(\vec{x}_0)$ are known.

Let us for simplicity assume that 

(i) $e_n$ are independent random variables (as for integrable systems) with
a step-like common distribution
\begin{equation}
d\mu (e) =\left \{ \begin{array}{rl} \frac{1}{2W}de,& \;\mbox{ if }|e|\le
  W\\ 0, & \;
\mbox{ otherwise} \end{array} \right .
\label{10}
\end{equation}
and (ii) $|\psi(\vec{x}_0)|^2=1$ (as for a rectangular billiard with periodic
boundary conditions).

Under these assumptions Eq.~(\ref{9}) takes the form
\begin{equation}
\sum_{j=1}^N \frac{1}{E-e_j}-\frac{1}{\lambda}=0
\label{11}
\end{equation}
where each term in the sum is an independent random variable.

Let us write the exact density of solutions of this equation  in the
following form
\begin{equation}
\rho (E) =\delta
(\sum_{j=1}^N\frac{1}{E-e_j}-\frac{1}{\lambda})\sum_{k=1}^N\frac{1}{(E-e_k)^2}.
\label{12}
\end{equation}
Representing the $\delta$-function as a Fourier integral one can express
this level density of through the characteristic function of Eq.~(\ref{11})
\begin{equation}
\rho (E) =\int_{-\infty}^{\infty}\frac{d\alpha}{2\pi}
\exp (i\alpha (\sum_{j=1}^N\frac{1}{E-e_j}-\frac{1}{\lambda}))\sum_{k=1}^N\frac{1}{(E-e_k)^2}.
\label{13}
\end{equation}
Because all $e_j$ are assumed to be independent random variables, the
correlation functions of $E$ can, in principle, be computed straightforwardly.

In particular, the 2-point correlation function
\begin{equation}
R_2(E_1,E_2)=<\rho (E_1) \rho (E_2)>
\label{14}
\end{equation}
can be expressed  in the following form
\begin{eqnarray}
R_2(E_1,E_2)&=&\int \frac{d\alpha_1 d\alpha_2}{4\pi^2}[ N
(f(\alpha_1,\alpha_2))^{N-1}g(\alpha_1,\alpha_2)
\label{15}\\
&+&
N(N-1)(f(\alpha_1,\alpha_2))^{N-2}\Psi_1(\alpha_1,\alpha_2)
\Psi_2(\alpha_1,\alpha_2)]e^{-i(\alpha_1+\alpha_2)/\lambda},
\nonumber
\end{eqnarray}
where
\begin{eqnarray}
f(\alpha_1,\alpha_2)&=&\int d\mu(e)\exp (i\frac{\alpha_1}{E_1-e}+
i\frac{\alpha_2}{E_2-e}),\nonumber\\
g(\alpha_1,\alpha_2)&=&\int d\mu(e)\exp (i\frac{\alpha_1}{E_1-e}+
i\frac{\alpha_2}{E_2-e})\frac{1}{(E_1-e)^2(E_2-e)^2},\nonumber\\
\Psi_i(\alpha_1,\alpha_2)&=&\int d\mu(e)\exp (i\frac{\alpha_1}{E_1-e}+
i\frac{\alpha_2}{E_2-e})\frac{1}{(E_i-e)^2}.\nonumber
\end{eqnarray}
When $N\rightarrow \infty$ the direct (but tedious) calculation of these
integrals gives \cite{9}
\begin{equation}
R_2(E_1,E_2)=\bar{\rho}^2r_2(\epsilon),
\label{16}
\end{equation}
where
\begin{eqnarray}
&&r_2(\epsilon)=1+
\label{17}\\
&&+\int_0^{\infty} d\alpha_1 \int_0^{\infty} d\alpha_2
(J_0^2(2\sqrt{\alpha_1\alpha_2}\ )+J_1^2(2\sqrt{\alpha_1\alpha_2}\ ))e^{-2\pi
  \epsilon J(\alpha_1,\alpha_2) +2i(\alpha_1+\alpha_2)},
\nonumber
\end{eqnarray}
and
\begin{eqnarray}
&&J(\alpha_1,\alpha_2)=\frac{1}{2}(\alpha_1+\alpha_2)+\frac{i}{2\pi \lambda_r}
(\alpha_1-\alpha_2)\nonumber\\
&&+ie^{i(\alpha_1+\alpha_2)}
[(\alpha_1-\alpha_2)G(\alpha_1,\alpha_2)
-i\alpha_1 J_0(2\sqrt{\alpha_1\alpha_2}\ )-
  \sqrt{\alpha_1\alpha_2}J_1(2\sqrt{\alpha_1\alpha_2}\ )],
\nonumber  
\end{eqnarray}
$$
G(\alpha_1,\alpha_2)=e^{i\alpha_2}\int_{\alpha_1}^{\infty}
J_0(2\sqrt{\alpha_2 t}\ )e^{it}dt,
$$
$$
\frac{1}{\lambda_r}=\frac{1}{\bar{\rho}\lambda}+\ln \frac{W-E}{W+E},\;
\epsilon =\bar{\rho}(E_1-E_2),\; \bar{\rho}=\frac{N}{2W}.
$$
This exact formula permits, in particular, to find the limiting behavior of the 2-point
correlation function. 

When $\epsilon \rightarrow 0$ 
\begin{equation}
r_2(\epsilon)\rightarrow A\epsilon 
\label{18}
\end{equation}
where
\begin{equation}
A=3\pi^2\int_0^{\infty} x
J_0((3+\delta)x)J_0^3(x)dx=\frac{\pi\sqrt{3}}{2}\approx 2.72
\label{19}
\end{equation}
is independent of the coupling constant  $\lambda$, and differs from the GOE
prediction ($r_2(\epsilon)\rightarrow (\pi^2/6) \epsilon \approx 1.64 \epsilon$).

When $\epsilon \rightarrow \infty$
\begin{equation}
r_2(\epsilon)\rightarrow 1+\frac{2}{\epsilon^2(\pi^2+1/\lambda_r^2)}.
\label{20}
\end{equation}
For GOE $r_2(\epsilon)\rightarrow 1-1/(\pi \epsilon)^2.$

These calculations give rise to different generalizations.

(i) The spectral statistics can also be computed for models where the
residues, $|\psi(\vec{x}_0)|^2$ are either fixed quantity (different from 1)
or independent random variables. In particular, for the rectangular billiard
with the
Dirichlet boundary conditions the introduction of the potential (\ref{7})
(when coordinates of the scatterer are non-commensurable with the sides)
leads to the following behavior of the 2-point correlation function at
small $\epsilon$
\begin{equation}
r_2(\epsilon)\rightarrow \frac{\epsilon}{8\pi^3}\ln^4 (\epsilon).
\label{21}
\end{equation}

(ii) The spectral statistics of the Bohr-Mottelson model \cite{10} which
describes the interaction of one level with all others can be
calculated by a similar method.

(iii) The asymptotic behavior of the 2-point correlation function 
(\ref{18})-(\ref{20}) can be derived without the knowledge of the exact
solution (\ref{17}) even for more general cases. 

(iv) In  \cite{10b} it was
demonstrated that in the summation over periodic orbits of integrable systems
there exists hidden saddle points whose contributions give correct
off-diagonal terms in the product of semiclassical terms like that in 
(\ref{2c}). 
This result permits to construct a specific
perturbation theory by which one can compute successive terms of 
the expansion at large $\epsilon$ of the 2-point correlation function 
for integrable systems with finite diffraction coefficient  by semiclassical
methods.

(v) Using the results of \cite{10b} it is possible to obtain an analog of
trace formulas for composite operators (build from a product of the Green
functions (\ref{1})), e.g. for conductance fluctuations
\cite{10c}.

\section{Pseudo-integrable billiards}

An interesting and important class of diffractive systems is 2-dimensional
polygonal billiards with angles, $\alpha_i$, equal rational multiples of $\pi$
\begin{equation}
\alpha_i=\frac{m_i}{n_i}\pi
\label{22}
\end{equation}
where $m_i$ and $n_i$ are coprime integers. 
These models (called pseudo-integrable billiards \cite{11, 12}) have a
characteristic property that their classical trajectories belong to a
2-dimensional surface with the genus 
\begin{equation}
g=1+\frac{N}{2}\sum_i\frac{m_i-1}{n_i}
\label{23}
\end{equation}
where $N$ is the least common multiplier of $n_i$.

Classical trajectories in these billiards are not defined after they hit a
corner  (\ref{22}) with  $m_i\ne 1$. These singular angles play the role of
diffraction centers and the quantum diffraction coefficient can be read off
from the exact Sommerfeld solution near a wedge with angle $\alpha$ \cite{13}
\begin{equation}
D(\theta_f, \theta_i)=\frac{2}{\gamma}\sin \frac{\pi}{\gamma}
\left [\frac{1}{\cos\frac{\pi}{\gamma}-\cos \frac{\theta_f+\theta_i}{\gamma}}-
\frac{1}{\cos\frac{\pi}{\gamma}-\cos \frac{\theta_f-\theta_i}{\gamma}} \right ]
\label{24}
\end{equation}
where $\gamma=\alpha/\pi$.

An important difference between  these models and the ones considered in the
previous Section is the fact that the diffraction coefficient (\ref{24}) 
diverges in certain directions
(called optical boundaries) to compensate the discontinuous behavior of
classical trajectories. Consequently, diffractive orbits lying on optical
boundaries cannot be described by Eq.~(\ref{3}) and, first of all, a
modification of the trace formulas is required \cite{14} which can  
conveniently be 
done  by using the Kirchoff approximation \cite{13} valid in a
vicinity of optical boundaries \cite{14}. The basis of this approximation is the
representation of the free (2-dimensional) Green function,
$G(\vec{x},\vec{x}\ ')$ as an integral over a line separating points $\vec{x}$ 
and $\vec{x}\ '$
\begin{equation}
G(\vec{x},\vec{x}\ ')=\int [G(\vec{q},\vec{x}\ ')
\frac{\partial}{\partial \vec{n}_q}G(\vec{x},\vec{q}\ )-
G(\vec{x},\vec{q}\ )
\frac{\partial}{\partial \vec{n}_q}G(\vec{q},\vec{x}\ ')]d\vec{q}
\label{25}
\end{equation}
where $\vec{n}_q$ is the normal to the line (parameterized by $\vec{q}\ $).

This formula is exact provided the integration is performed over the whole
(infinite) line. The existence of the wedge reduces the integration to a
semi-infinite region thus producing the Kirchoff approximation to the
diffraction problem. The detailed calculations of many particular types of
diffractive orbits in pseudo-integrable billiards have been performed in
Ref.~\cite{14}. Here we present only the contribution to the density of
states from diffractive
orbits lying on the boundary of multiple repetitions of a primitive periodic
orbit of length $l_0$
\begin{equation}
d_{l_0}(E)=-\frac{l_0}{8\pi k} \left (\sum_{q=1}^{\infty}
\frac{1}{q^{1/2}}e^{iqkl_0}\right )^2 +c.c.
\label{26}
\end{equation}
This contribution with fixed $l_0$ and $k\rightarrow \infty$ is smaller
that periodic orbit contribution (\ref{2c}) but bigger that the contribution
of diffractive orbits with finite diffraction coefficient (\ref{3}).

The diffraction coefficient (\ref{24})  can be written near an optical
boundary as a pole term plus a finite part, $D^{reg}$. Eq.~(\ref{26})
corresponds to the contribution from the pole term. It is also possible to
find analytically contributions from the interference of $D^{reg}$ with the
pole term
\begin{equation}
d_{l_0}'(E)=-\frac{1}{2\pi i}\frac{\partial}{\partial E}
\ln \left ( 1-\frac{D^{reg}}{\sqrt{8\pi k l_0}}
\sum_{q=1}^{\infty}\frac{1}{q^{3/2}}e^{iq k l_0 -3\pi i/4 -i\pi \nu_d/2}
\right ) +c.c.
\label{27}
\end{equation}
To derive these results it was necessary to compute analytically certain
multi-dimensional integrals. In particular, it has been shown in \cite{14} that 
\begin{eqnarray}
&&\int_0^{\infty}\ldots \int_0^{\infty} dx_1\ldots dx_n e^{-\Phi(\vec{x})}=
\nonumber\\
&&=\frac{1}{n+1}\int_{-\infty}^{\infty}\ldots \int_{-\infty}^{\infty} 
dx_1\ldots dx_n e^{-\Phi(\vec{x})}=\frac{\pi^{n/2}}{(n+1)^{3/2}}
\label{28}
\end{eqnarray}
where $\Phi(\vec{x})=x_1^2+(x_1-x_2)^2+\ldots +(x_{n-1}-x_n)^2+x_n^2$, and 
\begin{eqnarray}
&&\int_0^{\infty}dx[ 
\int_{-\infty}^{\infty}\ldots \int_{-\infty}^{\infty} 
dy_1\ldots dy_n e^{-\Psi(x,\vec{y})}-
\label{29} \\
&&
-\int_0^{\infty}\ldots \int_0^{\infty} dy_1\ldots dy_n e^{-\Psi(x,\vec{y})}]=
\frac{\pi^{(n-1)/2}}{4}\sum_{q=1}^{n-1}\frac{1}{\sqrt{q(n-q)}}
\nonumber
\end{eqnarray}
where $\Psi(x,\vec{y})=(x-y_1)^2+(y_1-y_2)^2+\ldots +(y_{n-1}-y_n)^2+(y_n-x)^2$.
The calculation of these integrals is based on the invariance of the
quadratic form $\Phi(\vec{x})$ under the action of a finite group generated
by reflections. 

Spectral statistics of pseudo-integrable billiards in the shape of the right
triangle with one angle equals $\pi/n$ for different $n$ has been
investigated numerically in Ref.~\cite{15} and it was demonstrated that the
energy level distribution for all $n$ (except the integrable cases
$n=3,4,6$) differs from both, the Poisson distribution typical for
integrable models and the random matrix distribution \cite{16} typical for
chaotic systems. It was observed numerically that the spectral statistics of
these billiards has the following characteristic properties

1. The 2-point correlation function $R_2(\epsilon)\rightarrow A \epsilon$
when $\epsilon \rightarrow 0$, i.e. there exist linear level repulsion as
for the Gaussian orthogonal ensemble of random matrices.

2. The nearest-neighbor distribution $p(s)\rightarrow e^{-\Lambda s}$ when
$s\rightarrow \infty$ as for the Poisson distribution.

3. The number variance  $\Sigma^2(L)\rightarrow \chi L$, 
when $L\rightarrow \infty$.

4. The distribution of energy levels do not change with increasing of energy 
(up to 30000 levels).

5. With increasing $n$ distributions stabilize quite far from random matrix
predictions.

6. For $n=5,8,10,12$ the spectral statistics is close (better than $10^{-2}$)
to the so-called semi-Poisson distribution \cite{17} which has the following
correlation functions
\begin{equation}
R_2(\epsilon)=1-e^{-4\epsilon},\;\; p(s)=4se^{-2s},\;\;\chi=\frac{1}{2}.
\label{30}
\end{equation}
Even the next-to-nearest distribution for these billiards agrees quite well
 with the prediction of the semi-Poisson model
\begin{equation}
p(2,s)=\frac{8}{3}s^3e^{-2s}.
\label{31}
\end{equation}

The first step of analytical investigation of spectral statistics for these
models is the calculation of the 2-point correlation form-factor in the
diagonal approximation \cite{18}. At small $\tau$ 
it is necessary to take into account only the contribution from periodic
orbits which in these cases is the same as for integrable billiards \cite{18}
\begin{equation}
K(\tau) = \frac{1}{2\pi^2 \bar{d}}\sum_p \frac{A_p}{l_p}\delta (l-4\pi k
\bar{d} \tau).
\label{32}
\end{equation}
The summation here is performed over all periodic orbits. $l_p$ is the
periodic orbit length, $A_p$ is the area occupied by primitive periodic 
orbit family. 

The billiard in the shape of the right triangle with one angle equals
$\pi/n$ belongs to the so-called Veech polygons \cite{19} for which there
exists a hidden group structure which permits the explicit calculation of
the number of periodic orbits and surfaces occupied by them. By generalizing
arguments of \cite{19} it is possible to prove \cite{19b} that for these triangular
billiards  
\begin{equation}
K(0)=\frac{n+\epsilon(n)}{3(n-2)}
\label{33}
\end{equation}
where $\epsilon (n)=0$ for $n$ odd, $\epsilon (n)=2$ for $n$ even but not
divisible by 3, and $\epsilon (n)=6$ for $n$ even and divisible by 3. 
These values of $K(0)$ are different from any known
distributions. The existence of non-zero $K(0)$ leads in particular to the
linear grows of the number variance, $\Sigma^2(L)\rightarrow K(0)L$ when
$L\rightarrow \infty$. 

The contributions from diffractive and non-diagonal terms are more difficult
to find. The main problem is the divergent character of the diffraction
coefficient which leads to the existence of terms as in Eq,~(\ref{26}) which
grow quickly with $l$ and cannot be treated in the diagonal
approximation. Such terms should be cancelled by other contributions and this
cancellation is a delicate procedure.

\section{Billiards with Aharonov-Bohm flux line}

Another interesting diffractive model close to the above polygonal billiards
is a rectangular billiard with the Aharonov-Bohm flux line \cite{20}. This
model is defined by the Schr\"odinger equation
\begin{equation}
[E+(\partial_{\mu}-iA_{\mu})^2]\Psi =0
\label{34}
\end{equation}
with the vector potential of a flux line $A_{\phi}=\alpha/r$ and, say, the
Dirichlet boundary conditions on the boundary of the billiard.

The introduction of the flux line does not change classical
trajectories. The only difference is that any time the trajectory encircles
the flux line it is necessary to add the phase $2\pi \alpha$ to the
semiclassical expansion (\ref{1}). 

The exact quantum-mechanical solution of the flux line in the whole space
gives  the value of the diffraction coefficient for the scattering on the
flux line \cite{20,21}
\begin{equation}
D(\theta_f, \theta_i)=\frac{2\sin \pi \alpha}{\cos
  (\frac{\theta_f-\theta_i}{2})}e^{i(\theta_f-\theta_i)/2}.
\label{35}
\end{equation}
This diffraction coefficient diverges in the forward direction and, as for
polygonal billiards, the trace formula requires the careful study of
multiple forward diffraction. This can be done, as in Section 3, by using
the (generalized) Kirchoff approximation \cite{14}. For example, the
contribution to the trace formula from the simplest diffractive orbit 
parallel to the base of the rectangle  (with 
2 points of forward diffraction) takes the form
\begin{equation}
d(E)=-\frac{2\sqrt{l_0(a-l_0})}{\pi^2 k}\sin^2 \pi \alpha \cos (2ka)  
\label{36}
\end{equation}
where $a$ is the base of the rectangle and $l_0$ is the distance of the flux
line from the rectangle side. Note that this expression has the same
dependence on momentum and periodic orbit length as Eq.~(\ref{26}).

For rectangular billiard with the flux line it is also possible to compute
the value of the 2-point correlation function at small $\tau$ \cite{19b}
When the ratios $x_0/a$ and $y_0/b$ of coordinates of the flux line
to the corresponding rectangle sides are non-commensurable irrational numbers
\begin{equation}
K(0)=1-3\bar{\alpha} +4\bar{\alpha}^2
\label{37}
\end{equation}
where $\bar{\alpha}$ is the fractional part of
$\alpha$, $0\le \bar{\alpha}\le 1/2$ and it is symmetric with respect to
$\bar{\alpha}=1/2$ when $1/2\le \bar{\alpha}\le 1$.

The numerically computed spectral statistics for this model shows
considerable deviations from any known distributions.

\section{Chaotic systems with diffractive center}

Consider now a chaotic system perturbed by a point-like scatterer. New
energy levels, as before, should be computed from Eq.~(\ref{9}). The only
difference with the case considered in Section 2 is that here we shall
assume that old (non-perturbed) energy levels, $e_n$, are distributed as
eigenvalues of one of the standard random matrix ensembles \cite{16}
\begin{equation}
P(\{e_k\})\propto \prod_{i<j}|e_i-e_j|^{\beta}
\label{39}
\end{equation}
where $\beta=1,2,4$ for, respectively, GOE, GUE, and GSE cases. (We
ignore here the one-body potential needed for the confinement of
eigenvalues. One can e.g. assume that the eigenvalues are lying on a large 
radius circle.)

Within the random matrix theory the distribution of $v_k=|\psi(\vec{x_0})|$ is
independent of the eigenvalues and is given by \cite{16}
\begin{equation}
P(\{v_k\}) =\prod_{i}(\frac{\beta N}{2\pi})^{1-\beta/2} \exp
  (-\frac{\beta}{2}Nv_i^2).
\label{40}
\end{equation}
The knowledge of the statistical distributions of the poles and the residues of
Eq.~(\ref{9}) permits the calculation of the distribution of new energy
levels. The first step has been done in Ref.~\cite{22} where the joint
distribution of the new, $E_j$, and the old, $e_i$, levels has been computed
\begin{equation}
P(\{E_j\},\{e_k\})\propto \frac{\prod_{i<j}(e_i-e_j)(E_i-E_j)}
{\prod_{i,j}|e_i-E_j|^{1-\beta/2}}\exp (-\rho \sum_i (E_i-e_i)),
\label{41}
\end{equation}
where $\rho=\beta/(2\lambda N)$ and, due to the positivity of $|v_k|^2$,
$e_i\le E_i\le e_{i+1}$. Here we assume that $\rho>0$ and energy levels are
ordered, $e_1\le e_2\le \ldots \le e_N$.

The resulting distribution of the new eigenvalues is defined by  the
expression
\begin{equation}
P(\{E_j\})=\int_{-\infty}^{E_1}de_1 \int_{E_1}^{E_2}de_2\ldots
\int_{E_{N-1}}^{E_N}de_NP(\{E_j\},\{e_k\}).
\label{42}
\end{equation}
In Ref.~\cite{23} it was demonstrated that these integrals can be computed
and the distribution $P(\{E_j\})$  has exactly the same form as
the distribution of non-perturbed eigenvalues given by Eq.~(\ref{31})
\begin{equation}
P(\{E_j\}) \propto \prod_{i<j}|E_i-E_j|^{\beta}.
\label{43}
\end{equation}
This result is not surprising. The random matrix theory distribution is
often considered as the result of the action of large number of small-size
scatterers. Consequently, the addition of one more center of scattering should not
change the spectral statistics. 

In Ref.~\cite{24} the contribution to the 2-point correlation form-factor
from diffractive orbits in the diagonal approximation has been computed for
f-dimen\-sio\-nal chaotic systems 
\begin{equation}
K_d(\tau)=\frac{\tau^2}{8\beta \pi^2}(\frac{k}{2\pi})^{2f-4}\int
|D(\vec{n},\vec{n}')|^2dO_{\vec{n}}dO_{\vec{n}'}.
\label{44}
\end{equation}
According to the above arguments this additional contribution should be
removed by other (non-diagonal) terms. In Ref.~\cite{23} it has been
demonstrated that it is the interference between diffractive orbits in the forward
direction and periodic orbits close to the diffraction center that exactly cancel
this term. In the derivation of this result two main ingredients were
important. First, the uniformity principle for periodic orbits of chaotic
systems, in particular 
\begin{equation}
\sum_p\frac{\chi(\vec{q}_p,\vec{p}_p)}{|\det (M_p-1)|}\delta (T-T_p)=
\frac{1}{\Sigma}\int \chi (\vec{q}, \vec{p}\ ) d^{f-1}q d^{f-1}p 
\label{45}
\end{equation}
where $\chi (\vec{q}, \vec{p}\ )$ is a test function defined on a Poincar\'e
surface of section $(\vec{q}, \vec{p}\ )$. $(\vec{q}_p, \vec{p}_p)$ are
coordinates of  points  of intersections of a periodic orbit with the surface of
section. $T_p$ is the periodic orbit period and 
$\Sigma=\int d^{f}q d^{f}p \delta (E-H(q,p))$ is the phase-space volume of
the constant energy surface.

The second important point is the optical theorem for the diffractive
coefficient which is a consequence of the unitarity of the scattering
$S$-matrix \cite{23}
\begin{equation}
\mbox{Im  }D(\vec{n},\vec{n})=-\frac{1}{8\pi}(\frac{k}{2\pi})^{f-2} \int
|D(\vec{n},\vec{n}')|^2dO_{\vec{n}'}.
\label{46}
\end{equation}
Using these relations one recovers \cite{23} the invariance of random matrix
results under the addition of  short-range scatterers from the semiclassical 
methods. One can easily check that ignoring the optical theorem can
change the spectral statistics.

\section{Conclusion}

In this paper a few typical examples of diffractive models has been
considered. An integrable model with a small-size scatterer with a constant
diffraction coefficient is the simplest and the most investigated case.
Under the assumption that the unperturbed system obeys the Poisson statistics
it is possible  compute rigorously the spectral statistics of this model. The
main result is that adding a diffractive center changes completely the
spectral statistics. The resulting statistics is characterized by level
repulsion and depends on the value of the diffractive coefficient. Models with
finite diffraction coefficient permit the construction of a specific
perturbation theory which allows the term-by term computation of expansion
of 2-point correlation form factor $K(\tau)$ (and other correlation
functions as well)  in powers of $\tau$. The models such as
pseudo-integrable polygonal billiards and rectangular billiards with a flux
line are another type of diffractive models characterized by divergence of
the diffraction coefficient in certain directions. In this case 
semiclassical contribution of
diffractive orbits to the trace formula differs from the case 
of finite diffraction coefficient 
and the spectral statistics of these models characterized, first of
all, by non-standard value of the 2-point correlation form factor at small
$\tau$, $K(0)<1$, and consequently the linear growths of the number variance.
For higher order terms of the expansion of correlation functions, 
the cancellation of rapidly growing terms (as in (\ref{26})) is not yet fully 
understood, and calculations are in progress. A simple example of a such
cancellation is provided by chaotic systems with a point-like scatterer.
Though it is evident (and can rigorously be proved) that when the spectral
statistics of unperturbed chaotic system is described by one of standard
random matrix ensembles, the addition of a diffraction center cannot change
the statistics, in semiclassical approach this invariance requires a
compensation between diagonal and off-diagonal terms. The important point is
that this strong cancellation is a general phenomenon connected mostly with
the unitarity of quantum mechanical scattering.

Diffractive systems are an interesting and promising class of
quan\-tum-me\-cha\-ni\-cal models. Their properties, in general,  differ from 
standard expectations but often  are accessible to analytical calculations 
and they open new perspectives in the
investigation of semiclassical limit in quantum mechanics.

{\bf Acknowledgments} 

The author is very grateful to C. Schmit, N. Pavloff, P. Leboeuf, U.
Gerland, and O. Giraud, whose collaboration permits to obtain most of the results 
of this paper.

\end{document}